\documentclass{revtex4}
\usepackage{amsthm,amssymb,amsmath}				
\usepackage{graphicx,comment}
\usepackage{float}   
 \usepackage{xcolor}

\begin{document}

\title{KG- oscillators in Som-Raychaudhuri rotating cosmic string spacetime
in a mixed magnetic field}
\author{Omar Mustafa}
\email{omar.mustafa@emu.edu.tr}
\affiliation{Department of Physics, Eastern Mediterranean University, 99628, G. Magusa,
north Cyprus, Mersin 10 - Turkiye.}

\begin{abstract}
\textbf{Abstract:} We investigate Klein-Gordon (KG) oscillators in a G\"{o}%
del-type Som-Raychaudhuri spacetime in a mixed magnetic field (given by the
vector potential $A_{\mu }=\left( 0,0,A_{\varphi },0\right) $, with $%
A_{\varphi }=B_{1}r^{2}/2+B_{2}r$). The resulting KG equation takes a Schr%
\"{o}dinger-like form (with an oscillator plus a linear plus a Coulomb-like
interactions potential) that admits a solution in the form of biconfluent
Heun functions/series $H_{B}\left( \alpha ,\beta ,\gamma ,\delta ,z\right) $%
. The usual power series expansion of which is truncated to a polynomial of
\ order $n_{r}+1=n\geq 1$ through the usual condition $\gamma =2\left(
n_{r}+1\right) +\alpha $. However, we use the very recent recipe suggested by
Mustafa \cite{1.29} as an alternative parametric
condition/correlation. i.e., $\delta =-\beta \left( 2n_{r}+\alpha +3\right) $, to
facilitate conditional exact solvability of the problem. We discuss and
report the effects of the mixed magnetic field as well as the effects of the
G\"{o}del-type SR-spacetime background on the KG-oscillators' spectroscopic
structure. 


\end{abstract}

\maketitle

\section{Introduction}

Modern superstring theories predict cosmic strings as one-dimensional stable
configurations of matter that are formed, along with other topological
defects in spacetime, during cosmological phase transitions in the early
universe \cite{1.1,1.2,1.3,1.4,1.5}. Cosmic strings are of particular
interest since their gravitational fields play a crucial role in galaxy
formation \cite{1.4,1.5}. Their gravitational fields introduce intriguing
effects like, to mention a few, self-interacting particles \cite{1.6,1.7},
gravitational lensing \cite{1.8}, and high energy particles \cite%
{1.9,1.10,1.11}. Nevertheless, their gravitational lensing \cite{1.8} is
believed to be the most effective way for their detection. Cosmic strings
(static or rotating) are characterized, in $\hbar =c=1$ units, by the wedge
parameter $\alpha =1-4\tilde{\mu}G$, which is a measure of angle deficit
produced by the string, where $G$ is the gravitational Newton constant and $%
\tilde{\mu}$ is the linear mass density of the string. A spinning/rotating
cosmic string has, however, an additional characteristic represented by a
rotational/spinning parameter. The spacetime metric that describes the
structure generated by a rotating cosmic string is given by%
\begin{equation}
ds^{2}=-\left( A\,dt+B\,d\varphi \right) ^{2}+dr^{2}+C^{2}\,d\varphi
^{2}+dz^{2},  \label{1.1}
\end{equation}%
where $A$, $B$, and $C$ are functions of the radial coordinate $r$ only \cite%
{1.12,1.13,1.14}. Which would, with $A=1$, $B=\alpha \Omega r^{2}$, and $%
C=\alpha r$, yield a G\"{o}del-type \cite{1.15,1.16} Som-Raychaudhuri (SR) 
\cite{1.17,1.18}\ spacetime metric%
\begin{equation}
ds^{2}=-\left( \,dt+\alpha \Omega r^{2}\,d\varphi \right) ^{2}+dr^{2}+\alpha
^{2}r^{2}\,d\varphi ^{2}+dz^{2},  \label{1.2}
\end{equation}%
where $0<\alpha <1$ in general relativity, $\alpha =1$ corresponds to
Minkowski spacetime, and $\alpha >1$ is used in the geometric theory of
topological defects in condense matter physics. Moreover, further other
solutions on rotating cosmic strings are investigated in \cite%
{1.181,1.182,1.183,1.184}. 

In the current methodical proposal, however, we shall be interested in
Klein-Gordon (KG) oscillators in Som-Raychaudhuri rotating cosmic string
spacetime (\ref{1.2}) in a mixed magnetic field. The corresponding covariant
and contravariant metric tensors of which are given by

\begin{equation}
g_{\mu \nu }=\left( 
\begin{tabular}{cccc}
$-1$ & $0$ & $-\alpha \Omega r^{2}$ & $0$ \\ 
$0$ & $1$ & $0$ & $0$ \\ 
$-\alpha \Omega r^{2}$ & $0$ & $(\alpha ^{2}r^{2}-\alpha ^{2}\Omega
^{2}r^{4})$ & $0$ \\ 
$0$ & $0$ & $0$ & $1$%
\end{tabular}%
\right) ;\;\det \left( g_{\mu \nu }\right) =g=-\alpha ^{2}r^{2},\;g^{\mu \nu
}=\left( 
\begin{tabular}{cccc}
$\left( \Omega ^{2}r^{2}-1\right) $ & $0$ & $-\frac{\Omega }{\alpha }$ & $0$
\\ 
$0$ & $1$ & $0$ & $0$ \\ 
$-\frac{\Omega }{\alpha }$ & $0$ & $\frac{1}{\alpha ^{2}r^{2}}$ & $0$ \\ 
$0$ & $0$ & $0$ & $1$%
\end{tabular}%
\right) .  \label{1.3}
\end{equation}%
Yet, the metric element $g_{\varphi \varphi }>0$ would suggest an upper
limit for the radial coordinate so that $0\leq r<1/|\Omega |$. Moreover, we
shall consider a 4-vector potential $A_{\mu }=\left( 0,0,A_{\varphi
},0\right) $, with $A_{\varphi }=A_{1\varphi }+A_{2\varphi
}=B_{1}r^{2}/2+B_{2}r$, to yield a mixed magnetic field $\mathbf{B}=B_{z}%
\hat{z}$, so that $B_{z}^{2}=F_{\mu \nu }F^{\mu \nu }/2\Longrightarrow B_{z}=%
\frac{1}{\alpha }\left( B_{1}+B_{2}/r\right) $. Here, $B_{1}$ denotes the
strength of a uniform magnetic field and $B_{2}$ denotes the strength of a
non-uniform magnetic field. KG-particles in the G\"{o}del-type
Som-Raychaudhuri cosmic string spacetime background have been intensively
studied with a uniform or a non-uniform magnetic fields \cite%
{1.18,1.186,1.187,1.188,1.189,1.1810,1.1811,1.1812,1.1813,1.1814,1.19,1.20,1.21,1.22,1.23,1.24,1.25,1.26,1.27,1.28,1.29,1.30}%
, but never with the mixed magnetic field above. 

Owing to the fact that Schr\"{o}dinger, Dirac, and KG oscillators are
quantum mechanically of fundamental pedagogical interest, the study of their
spectroscopic structure under the effects of the gravitational fields,
introduced by different spacetime fabrics, should be of fundamental
pedagogical interest in quantum gravity (c.f., e.g., sample of references 
\cite%
{1.31,1.32,1.33,1.34,1.35,1.36,1.37,1.38,1.39,1.40,1.41,1.42,1.43,1.44,1.45,1.46}%
, and related references cited therein). We have very recently studied
KG-particles in a cosmic string rainbow gravity spacetime in the mixed
magnetic field \cite{1.185}. Therein, we have observed that the competition
between the two magnetic field strengths $B_{1}$ and $B_{2}$ has generated
energy levels crossing which, consequently, turned the spectra upside down.
That was the only mixed magnetic field scenario in the literature, to the
best of our knowledge. It would be, therefore, interesting to study the
gravitational effects of the G\"{o}del-type Som-Raychaudhuri cosmic string
spacetime on the KG-particles in the mixed magnetic field. Our motivation to
carry out the current study is clear, therefore.

The organization of this study is in order. In section 2, we start with
KG-particles in G\"{o}del-type SR-spacetime in a mixed magnetic field and
including the KG-oscillators. The resulting differential equation is a Schr%
\"{o}dinger-like (with an effective potential that includes an oscillator
plus a linear plus a Coulomb like interactions) and admits a solution in the
form of biconfluent Heun functions/series \cite{1.47,1.48}. A three terms
recursion relation is manifestly introduced and a conditionally exact
solution is reported. Such a conditionally exact solution reproduces the
usual truncation of the biconfluent Heun function/series $H_{B}\left( \breve{%
\alpha},\beta ,\gamma ,\delta ,z\right) $ to a polynomial of order $%
n_{r}+1=n\geq 1$ (through the condition $\gamma =2\left( n_{r}+1\right) +%
\breve{\alpha}$) and provides a parametric correlation that allows us to
find conditionally exact solutions. We report the effects of the mixed
magnetic field as well as that of the G\"{o}del-type SR-spacetime
background. Our concluding remarks are given in section 3.

\section{KG-particles in G\"{o}del-type SR-spacetime in a mixed magnetic
field}

The KG-particles are described by the KG-equation 
\begin{equation}
\left( \frac{1}{\sqrt{-g}}\tilde{D}_{\mu }\sqrt{-g}g^{\mu \nu }\tilde{D}%
_{\nu }\right) \,\Psi \left( t,r,\varphi ,z\right) =m_{\circ }^{2}\,\Psi
\left( t,r,\varphi ,z\right) ,  \label{2.1}
\end{equation}%
where $\tilde{D}_{\mu }=\partial _{\mu }-ieA_{\mu }+\mathcal{F}_{\mu }$ with 
$\mathcal{F}_{\mu }$ $\in 
\mathbb{R}
$. One should notice that $\mathcal{F}_{\mu }=\left( 0,\mathcal{F}%
_{r},0,0\right) $ is in a non-minimal coupling form, whereas the 4-vector
potential $A_{\mu }=\left( 0,0,A_{\varphi },0\right) $ is in the usual
minimal coupling form, and $m_{\circ }$ denotes the rest mass energy (i.e., $%
m_{\circ }\equiv m_{\circ }c^{2}$, with $\hbar =c=1$ units to be used
throughout). With 
\begin{equation}
\Psi \left( t,r,\varphi ,z\right) =R\left( r\right) e^{i\left( m\varphi
+kz-Et\right) },  \label{2.2}
\end{equation}%
would read%
\begin{equation}
\left\{ \frac{1}{r}\left( \partial _{r}+\mathcal{F}_{r}\right) \,r\,\left(
\partial _{r}-\mathcal{F}_{r}\right) -\frac{\left( m-eA_{\varphi }\right)
^{2}}{\alpha ^{2}r^{2}}-\Omega ^{2}E^{2}r^{2}+\frac{2\Omega EeA_{\varphi }}{%
\alpha }+\mathcal{E}^{2}\right\} R\left( r\right) =0,  \label{2.3}
\end{equation}%
where $m$ is the the magnetic quantum number $m=m_{\pm }=\pm |m|=0,\pm 1,\pm
2,\cdots $, and  
\begin{equation}
\mathcal{E}^{2}=E^{2}-\frac{2\Omega Em}{\alpha }-\left( m_{\circ
}^{2}+k^{2}\right) .  \label{2.31}
\end{equation}%
We now use $A_{\varphi }=B_{1}r^{2}/2+B_{2}r$ and incorporate the
KG-oscillators through the substitution $\mathcal{F}_{r}=\eta r$  \cite%
{1.31,1.32} to obtain%
\begin{equation}
\left\{ \partial _{r}^{2}+\frac{1}{r}\partial _{r}-\frac{\tilde{m}^{2}}{r^{2}%
}-\tilde{\Omega}^{2}r^{2}-Ar+\frac{2\tilde{m}\tilde{B}_{2}}{r}+\mathcal{%
\tilde{E}}^{2}\right\} R\left( r\right) =0,  \label{2.4}
\end{equation}%
with%
\begin{eqnarray}
\mathcal{\tilde{E}}^{2} &=&\mathcal{E}^{2}-2\eta -\tilde{B}_{2}^{2}+\tilde{m}%
\tilde{B}_{1},\;A=\tilde{B}_{1}\tilde{B}_{2}-2\Omega E\tilde{B}_{2},
\label{2.51} \\
\tilde{\Omega}^{2} &=&\Omega ^{2}E^{2}+\eta ^{2}+\frac{\tilde{B}_{1}^{2}}{4}%
-\Omega E\tilde{B}_{1},\;\tilde{m}=\frac{m}{\alpha },\;\tilde{B}_{j}=\frac{%
eB_{j}}{\alpha }.  \label{2.52}
\end{eqnarray}%
We may now substitute%
\begin{equation}
R\left( r\right) =U\left( r\right) \,\exp \left( -\frac{|\tilde{\Omega}|}{2}%
\left[ r+\frac{A}{2\tilde{\Omega}^{2}}\right] \right)   \label{2.6}
\end{equation}
in (\ref{2.4}) to obtain%
\begin{gather}
r^{2}U^{^{\prime \prime }}\left( r\right) -\left[ 2|\tilde{\Omega}|r^{3}+%
\frac{A}{|\tilde{\Omega}|}r^{2}-r\right] U^{\prime }\left( r\right) +\left[
\left( \mathcal{\tilde{E}}^{2}+\frac{A^{2}}{4\tilde{\Omega}^{2}}-2|\tilde{%
\Omega}|\right) r^{2}\right.   \notag \\
\left. +\left( 2\tilde{m}\tilde{B}_{2}-\frac{A}{2|\tilde{\Omega}|}\right) r-%
\tilde{m}^{2}\right] U\left( r\right) =0.  \label{2.7}
\end{gather}%
This equation is known to have a solution in the form of biconfluent Heun
functions so that%
\begin{equation}
U\left( r\right) =N_{1}\,r^{\left\vert \tilde{m}\right\vert }\,H_{B}\left( 
\breve{\alpha},\beta ,\gamma ,\delta ,z\right) +N_{2}\,r^{-\left\vert \tilde{%
m}\right\vert }\,H_{B}\left( -\breve{\alpha},\beta ,\gamma ,\delta ,z\right) 
\label{2.71}
\end{equation}%
where 
\begin{equation}
\breve{\alpha}=2\left\vert \tilde{m}\right\vert ,\,\beta =\frac{A}{|\tilde{%
\Omega}|^{3/2}},\,\gamma =\frac{\mathcal{\tilde{E}}^{2}}{|\tilde{\Omega}|}+%
\frac{A^{2}}{4|\tilde{\Omega}|^{3}},\,\delta =-\frac{4\tilde{m}\tilde{B}_{2}%
}{\sqrt{|\tilde{\Omega}|}},\,z=\sqrt{|\tilde{\Omega}|}r.  \label{2.72}
\end{equation}%
Obviously, we have to take $N_{2}=0$ to secure finiteness of the solution at 
$r=0$. Consequently, one may cast the solution as%
\begin{equation}
U\left( r\right) =N_{1}\,r^{\left\vert \tilde{m}\right\vert }\,H_{B}\left( 
\breve{\alpha},\beta ,\gamma ,\delta ,z\right) .  \label{2.73}
\end{equation}%
We now need to truncate the biconfluent Heun function to a polynomial of
order $n_{r}+1$. This is done when the condition $\gamma =2\left(
n_{r}+1\right) +\breve{\alpha}$ is sought \cite{1.47,1.48} to obtain%
\begin{equation}
\mathcal{\tilde{E}}^{2}=2|\tilde{\Omega}|\left( n_{r}+|\tilde{m}|+1\right) -%
\frac{A^{2}}{4\tilde{\Omega}^{2}}.  \label{2.74}
\end{equation}%
Moreover, the biconfluent Heun series could admit further truncation through
the assumption that the  $n_{r}+1$ coefficient in the series expansion is
a polynomial of degree $n_{r}$ in $\delta $, provided that $\delta $ is a
root of this polynomial which consequently cancels , the  $n_{r}+1$
subsequent coefficients and the series truncates to degree $n_{r}$ for $%
H_{B}\left( \breve{\alpha},\beta ,\gamma ,\delta ,z\right) $ \cite{1.47}. In
the current methodical proposal, however, we shall follow a new recipe that
manifestly introduces a clear correlation between the physical parameters
involved and truncates the series to a polynomial of order $n_{r}+1=n\geq 1$%
. This is done in the sequel.

We follow the power series expansion of the biconfluent Heun function as
usual so that%
\begin{equation}
U\left( r\right) =r^{\nu }\sum\limits_{j=0}^{\infty }C_{j}r^{j},  \label{2.8}
\end{equation}%
to obtain%
\begin{gather}
\sum\limits_{j=0}^{\infty }\left\{ C_{j+2}\left[ \left( j+\nu +2\right) ^{2}-%
\tilde{m}^{2}\right] +C_{j+1}\left[ 2\tilde{m}\tilde{B}_{2}-\frac{A}{2|%
\tilde{\Omega}|}\left( 2j+2\nu +3\right) \right] \right.   \notag \\
+\left. C_{j}\left[ \mathcal{\tilde{E}}^{2}+\frac{A^{2}}{4\tilde{\Omega}^{2}}%
-2|\tilde{\Omega}|\left( j+\nu +1\right) \right] \right\} r^{j+\nu
+2}+C_{0}\left( \nu ^{2}-\tilde{m}^{2}\right) r^{\nu }  \notag \\
+\left( C_{0}\left[ 2\tilde{m}\tilde{B}_{2}-\frac{A}{2|\tilde{\Omega}|}%
\left( 2\nu +1\right) \right] +C_{1}\left[ \left( \nu +1\right) ^{2}-\tilde{m%
}^{2}\right] \right) r^{\nu +1}=0  \label{2.9}
\end{gather}%
This necessarily implies that%
\begin{equation}
C_{0}\neq 0\Longrightarrow \nu ^{2}-\tilde{m}^{2}=0\Longrightarrow \nu =\pm |%
\tilde{m}|,  \label{2.91}
\end{equation}%
where $\nu =+|\tilde{m}|$ is the value to be adopted (otherwise, the wave
function is divergent at $r=0)$, and 
\begin{equation}
C_{1}=\left( \frac{A}{2|\tilde{\Omega}|}-\frac{2\tilde{m}\tilde{B}_{2}}{%
\left( 2|\tilde{m}|+1\right) }\right) C_{0}.  \label{2.92}
\end{equation}%
Consequently, we obtain a three terms recursion relation in the form of%
\begin{gather}
C_{j+2}\left[ \left( j+|\tilde{m}|+2\right) ^{2}-\tilde{m}^{2}\right]
+C_{j+1}\left[ 2\tilde{m}\tilde{B}_{2}-\frac{A}{2|\tilde{\Omega}|}\left(
2j+2|\tilde{m}|+3\right) \right]   \notag \\
+C_{j}\left[ \mathcal{\tilde{E}}^{2}+\frac{A^{2}}{4\tilde{\Omega}^{2}}-2|%
\tilde{\Omega}|\left( j+|\tilde{m}|+1\right) \right] =0,\,j\geq 0.
\label{2.93}
\end{gather}%
Which in turn gives for $j=0$%
\begin{equation}
C_{2}=\frac{C_{1}\left[ \frac{A}{2|\tilde{\Omega}|}\left( 2|\tilde{m}%
|+3\right) -2\tilde{m}\tilde{B}_{2}\right] +C_{0}\left[ 2|\tilde{\Omega}%
|\left( |\tilde{m}|+1\right) -\left( \mathcal{\tilde{E}}^{2}+\frac{A^{2}}{4%
\tilde{\Omega}^{2}}\right) \right] }{4\left( |\tilde{m}|+1\right) },
\label{2.94}
\end{equation}%
and so on we find the coefficients $C_{j}^{\prime }s$ for the power series (%
\ref{2.8}). However, finiteness and square integrability of the wave
function requires that the power series should be truncated into a
polynomial. To do so, we follow the recipe we used in \cite{1.29,1.46}. That
is, $\forall j=n_{r}$ we take $C_{n_{r}+2}=0$, $C_{n_{r}+1}\neq 0$, and $%
C_{n_{r}}\neq 0$. One should notice that the condition $C_{n_{r}+2}=0$ would
allow us to obtain a polynomial of order $n_{r}+1\geq 1$. However, we
further require that the coefficients of $C_{n_{r}+1}\neq 0$ and $%
C_{n_{r}}\neq 0$ to vanish identically to allow conditional exact
solvability of the problem at hand. Under such conditional exact
solvability, we obtain%
\begin{equation}
C_{n_{r}+1}\neq 0\Longrightarrow 2\tilde{m}\tilde{B}_{2}-\frac{A}{2|\tilde{%
\Omega}|}\left( 2n_{r}+2|\tilde{m}|+3\right) =0\Longrightarrow 2|\tilde{%
\Omega}|=\left\vert \frac{\tilde{B}_{1}-2\Omega E}{\tilde{m}}\right\vert
\left( \tilde{n}+\frac{1}{2}\right) ,  \label{2.95}
\end{equation}%
and 
\begin{equation}
C_{n_{r}}\neq 0\Longrightarrow \mathcal{\tilde{E}}^{2}+\frac{A^{2}}{4\tilde{%
\Omega}^{2}}-2|\tilde{\Omega}|\left( n_{r}+|\tilde{m}|+1\right)
=0\Longrightarrow \mathcal{\tilde{E}}^{2}=2|\tilde{\Omega}|\tilde{n}-\frac{%
A^{2}}{4\tilde{\Omega}^{2}}.  \label{2.96}
\end{equation}%
where $\tilde{n}=n_{r}+|\tilde{m}|+1$. At this point, one should notice that
whilst condition (\ref{2.95}) offers conditional exact solvability through a
parametric correlation, the second condition (\ref{2.96}) is in exact accord
with that (\ref{2.74}) and provides the KG-oscillators energies as%
\begin{equation}
E^{2}+\tilde{m}\left( \tilde{B}_{1}-2\Omega E\right) -\tilde{n}\left( \tilde{%
n}+\frac{1}{2}\right) \left\vert \frac{\tilde{B}_{1}-2\Omega E}{\tilde{m}}%
\right\vert -\mathcal{G}_{n_{r},m}=0,  \label{2.97}
\end{equation}%
with%
\begin{equation}
\mathcal{G}_{n_{r},m}=m_{\circ }^{2}+k^{2}+2\eta +\tilde{B}_{2}^{2}-\frac{%
\tilde{m}^{2}\tilde{B}_{2}^{2}}{\left( \tilde{n}+1/2\right) ^{2}}.
\label{2.98}
\end{equation}%
One would observe that this quadratic energy equation is unlikely to be
analytically solvable. However, to observe the effects of the uniform $B_{1}$
and the non-uniform $B_{2}$ magnetic fields, and vorticity $\Omega $, we
plot the KG-particles' and antiparticles' energies in (\ref{2.97}) in
figures 1 and 2. For some fixed values of $\left( \alpha ,\eta ,k,m_{\circ
}\right) =\left( 0.5,1,1,1\right) $ we plot in Fig.1 the energies $E$
against vorticity $\Omega $ so that 1(a) for $B_{1}=0$, $B_{2}=1$, 1(b) for $%
B_{1}=1$, $B_{2}=0$, and 1(c) for $B_{1}=1$, $B_{2}=1$. Where as, in Fig.2
we plot the energies $E$ against $B_{1}$, in 2(a) and 2(b), and against $%
B_{2}$, in 2(c) and 2(d), for vorticity $\Omega =\pm 1$. A common
characteristic of all such figures is that the symmetry of the energies
about $E=0$ is broken. We observe that, while the minima of $\left\vert
E_{\pm }\right\vert $ are located at $\Omega =0$ value when the uniform
magnetic field is switched off ($B_{1}=0$), they shift to $\left(
E_{+},\Omega _{+}\right) $ and $\left( E_{-},\Omega _{-}\right) $ quarters
of 1(b) and 1(c) for $B_{1}\neq 0$. This is due to the fact that we
effectively have%
\begin{equation}
\left\vert \tilde{\Omega}\right\vert =\sqrt{\left( \Omega E-\frac{\tilde{B}%
_{1}}{2}\right) ^{2}+\eta ^{2}}.  \label{2.98.1}
\end{equation}%
Where it is obvious that the term $\Omega E=\pm \left\vert \Omega
E\right\vert $ is competing with $\tilde{B}_{1}/2\geq 0$. That is, the first
term under the square root in (\ref{2.98.1}) takes the values 
\begin{equation}
\left( \Omega E-\frac{\tilde{B}_{1}}{2}\right) ^{2}=\left\{ 
\begin{array}{c}
\left( \Omega _{\pm }E_{\pm }-\frac{\tilde{B}_{1}}{2}\right) ^{2}=\left(
\left\vert \Omega \right\vert \left\vert E\right\vert -\frac{\tilde{B}_{1}}{2%
}\right) ^{2},\text{ for }\Omega E=\Omega _{\pm }E_{\pm } \\ 
\left( \Omega _{\mp }E_{\pm }-\frac{\tilde{B}_{1}}{2}\right) ^{2}=\left(
-\left\vert \Omega \right\vert \left\vert E\right\vert -\frac{\tilde{B}_{1}}{%
2}\right) ^{2},\text{ for }\Omega E=\Omega _{\mp }E_{\pm }%
\end{array}%
\right. ,  \label{2.98.2}
\end{equation}%
\begin{figure}[ht!]  
\centering
\includegraphics[width=0.3\textwidth]{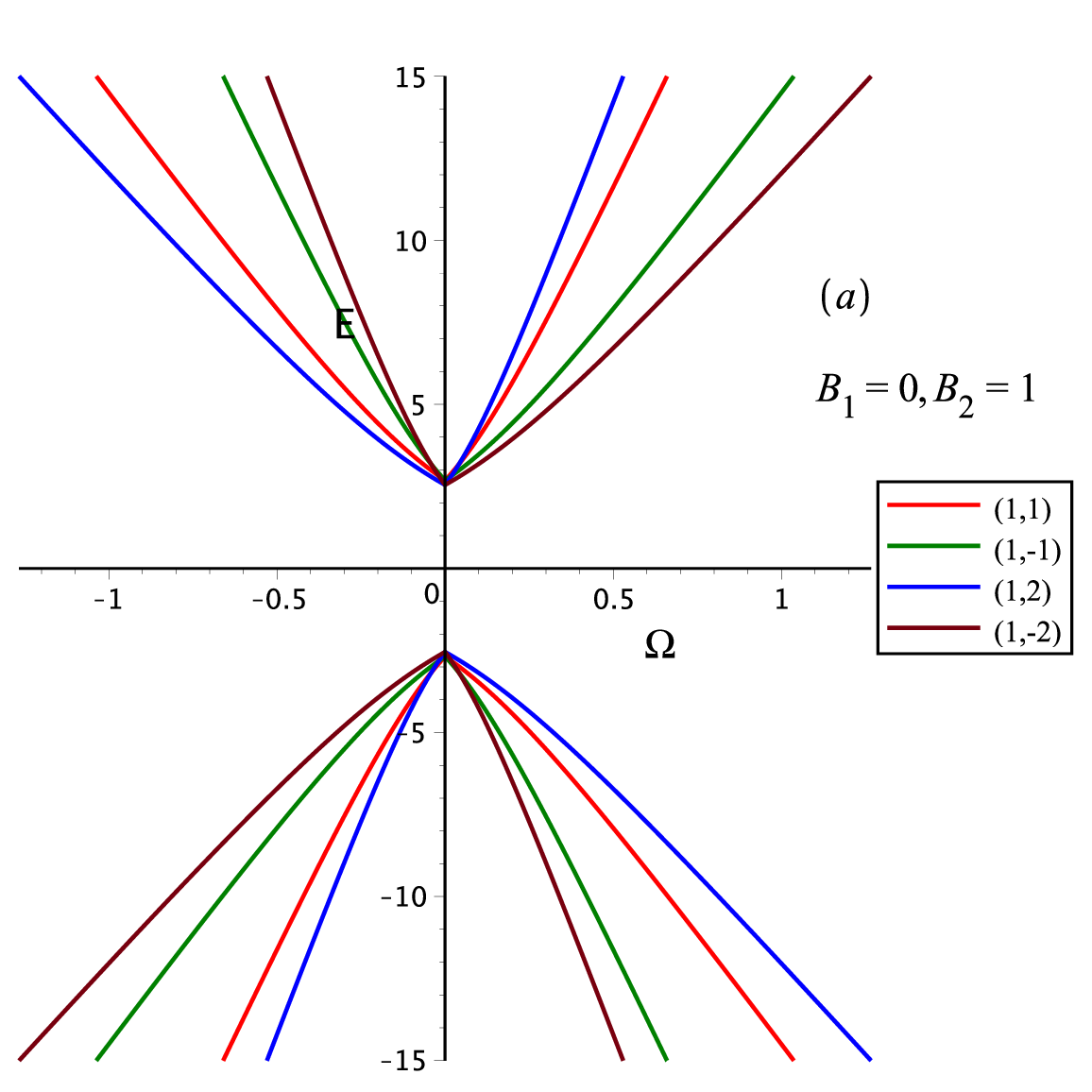}
\includegraphics[width=0.3\textwidth]{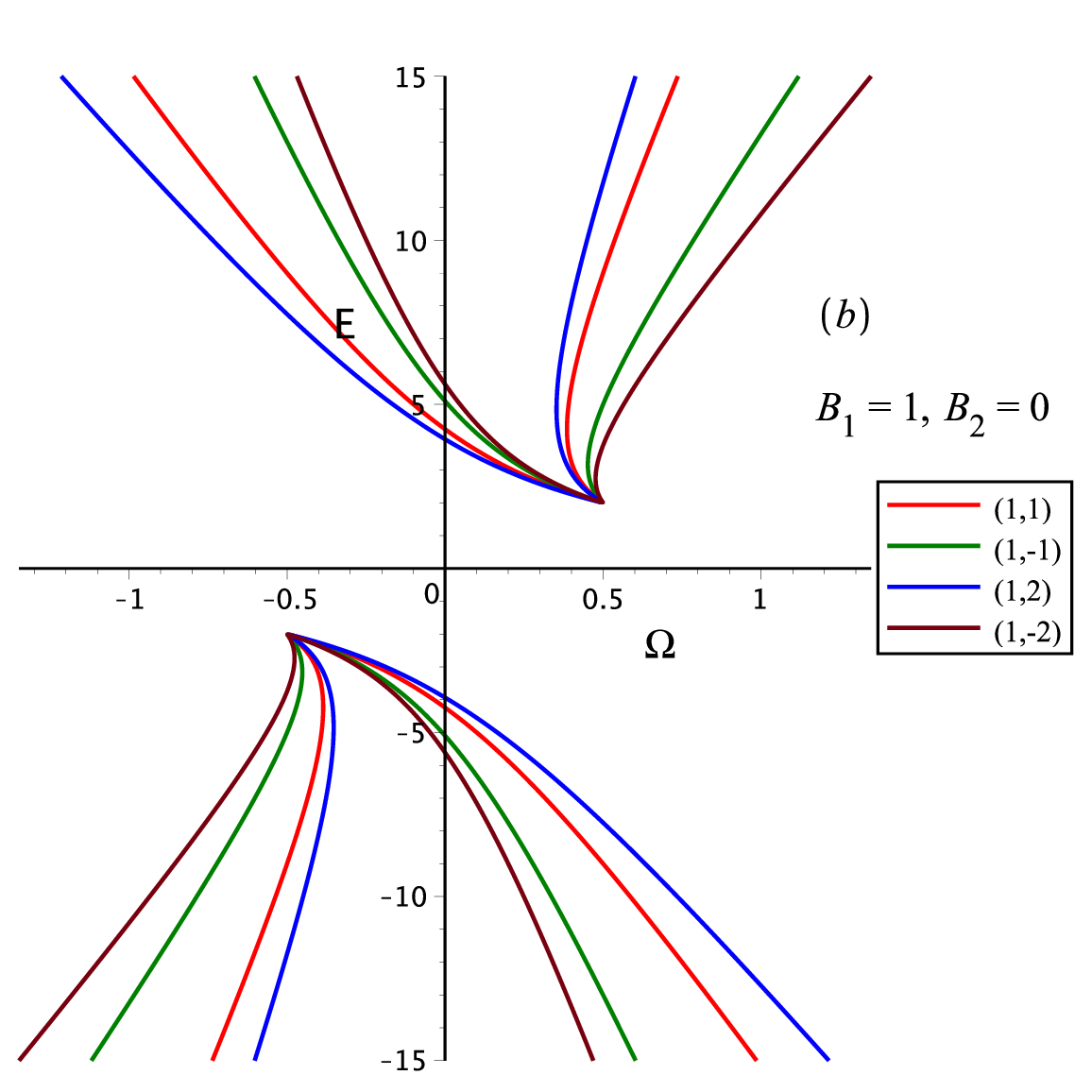}
\includegraphics[width=0.3\textwidth]{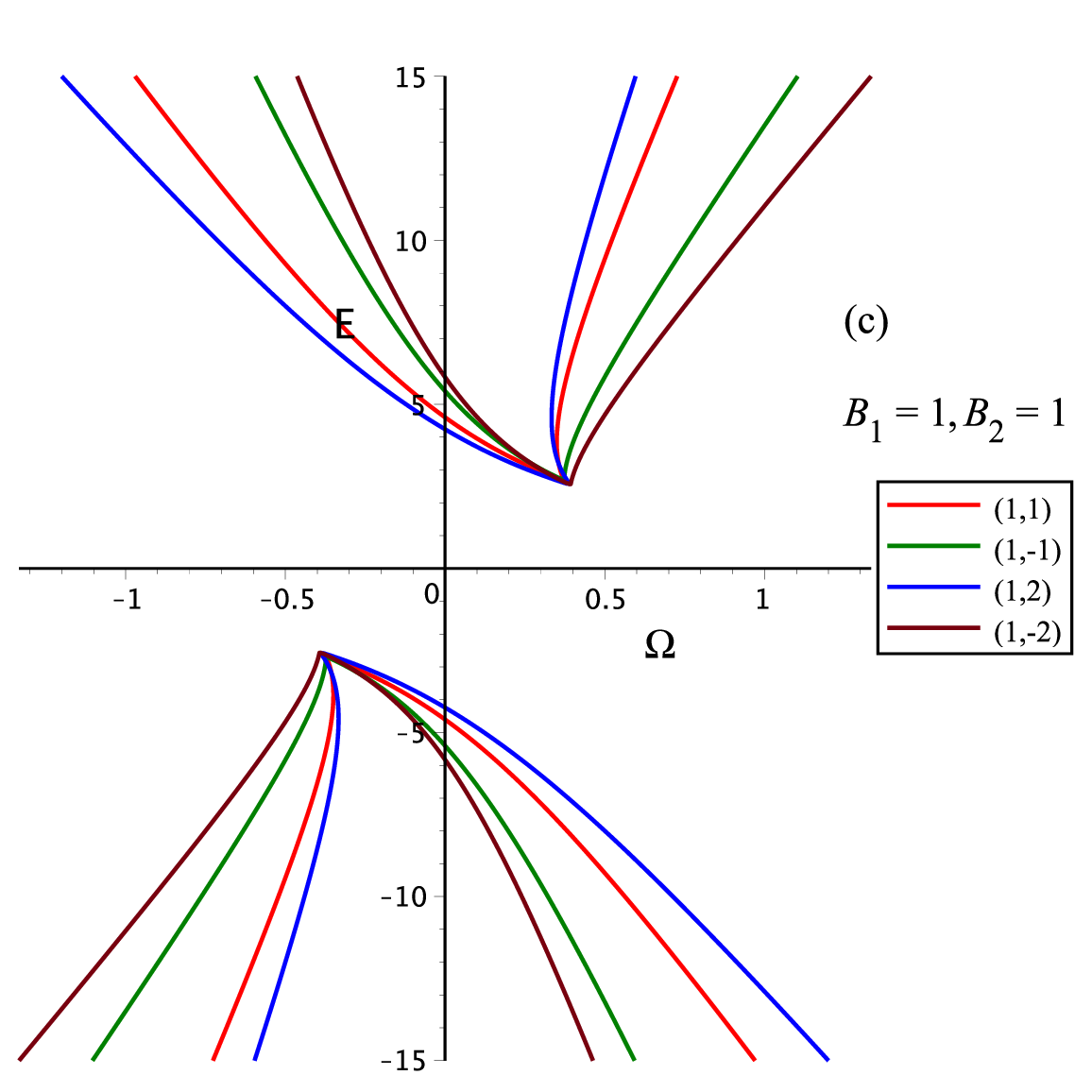}
\caption{\small 
{ The energy levels against the vorticity $\Omega $ for $%
\left( n_{r},m\right) $ states with $n_{r}=1$ and $m=\pm 1,\pm 2$ given by
Eq. (\ref{2.97}), at $\alpha =0.5$, and $m_{\circ }=1=\eta =k$, so that
Fig.1(a) for $B_{1}=0$ and $B_{2}=1$, 1(b) for $B_{1}=1$ and $B_{2}=0$, and
1(c) for $B_{1}=1$ and $B_{2}=1$.}}
\label{fig1}
\end{figure}%
and consequently suggest that 
\begin{equation}
\left( \Omega _{\pm }E_{\pm }-\frac{\tilde{B}_{1}}{2}\right) ^{2}<\left(
\Omega _{\mp }E_{\pm }-\frac{\tilde{B}_{1}}{2}\right) ^{2}.  \label{2.98.3}
\end{equation}
Obviously, therefore, the competition between an effective energy dependent
vorticity (i.e., $\grave{\Omega}\left( E\right) =\Omega E=\pm \left\vert
\Omega \right\vert \left\vert E\right\vert $) and the uniform magnetic field
strength $\tilde{B}_{1}$ in (\ref{2.98.1}) plays a crucial role in shaping
the spectroscopic structure of the KG-oscillators in Som-Raychaudhuri
rotating cosmic string spacetime in a mixed magnetic field. This would
explain the similar behaviours of the curves in the first and third quarters
for $\left( E_{+},\Omega _{+}\right) $ and $\left( E_{-},\Omega _{-}\right) $%
, respectively, as well as the similar behaviours of the curves in the
second and fourth quarters for $\left( E_{-},\Omega _{+}\right) $ and $%
\left( E_{+},\Omega _{-}\right) $, respectively. of Fig.s 1(b) and 1(c).
Moreover, fixing the values of $\Omega $ so that $\Omega =\pm 1$, we observe
that in Fig. 2(a) $\left\vert E_{+}\right\vert $ decreases whereas $%
\left\vert E_{-}\right\vert $ increases as as $B_{1}$ increases (for a fixed $%
B_{2}=1$) for the vorticity $\Omega =+1$, Whereas, in Fig. 2(b), $\left\vert
E_{+}\right\vert $ increases whereas $\left\vert E_{-}\right\vert $
decreases as as $B_{1}$ increases (for a fixed $B_{2}=1$) for the vorticity $%
\Omega =-1$. In Fig.s 2(c) and 2(d) we observe that both $\left\vert
E_{+}\right\vert $ and $\left\vert E_{-}\right\vert $ increases slowly with
increasing $B_{2}$ for a fixed $B_{1}=1$ value. Yet, in all the figures
reported, we clearly observe that the symmetry of the energies about $E=0$
value is broken mainly because of the effect of the uniform magnetic field $%
B_{1}$ (which is obvious in (\ref{2.98.3})). This is to be shown in the
sequel.
\begin{figure}[ht!]  
\centering
\includegraphics[width=0.35\textwidth]{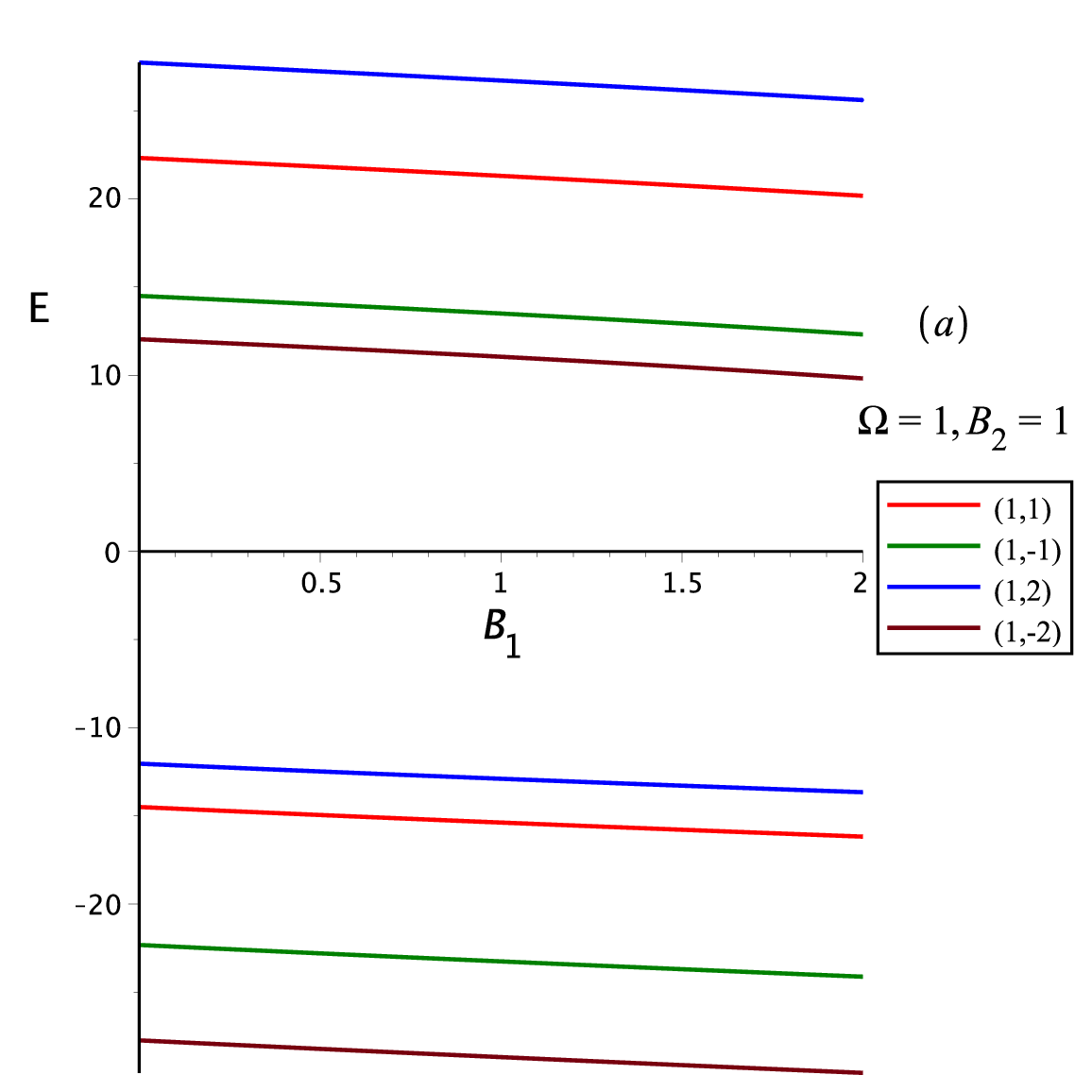}
\includegraphics[width=0.35\textwidth]{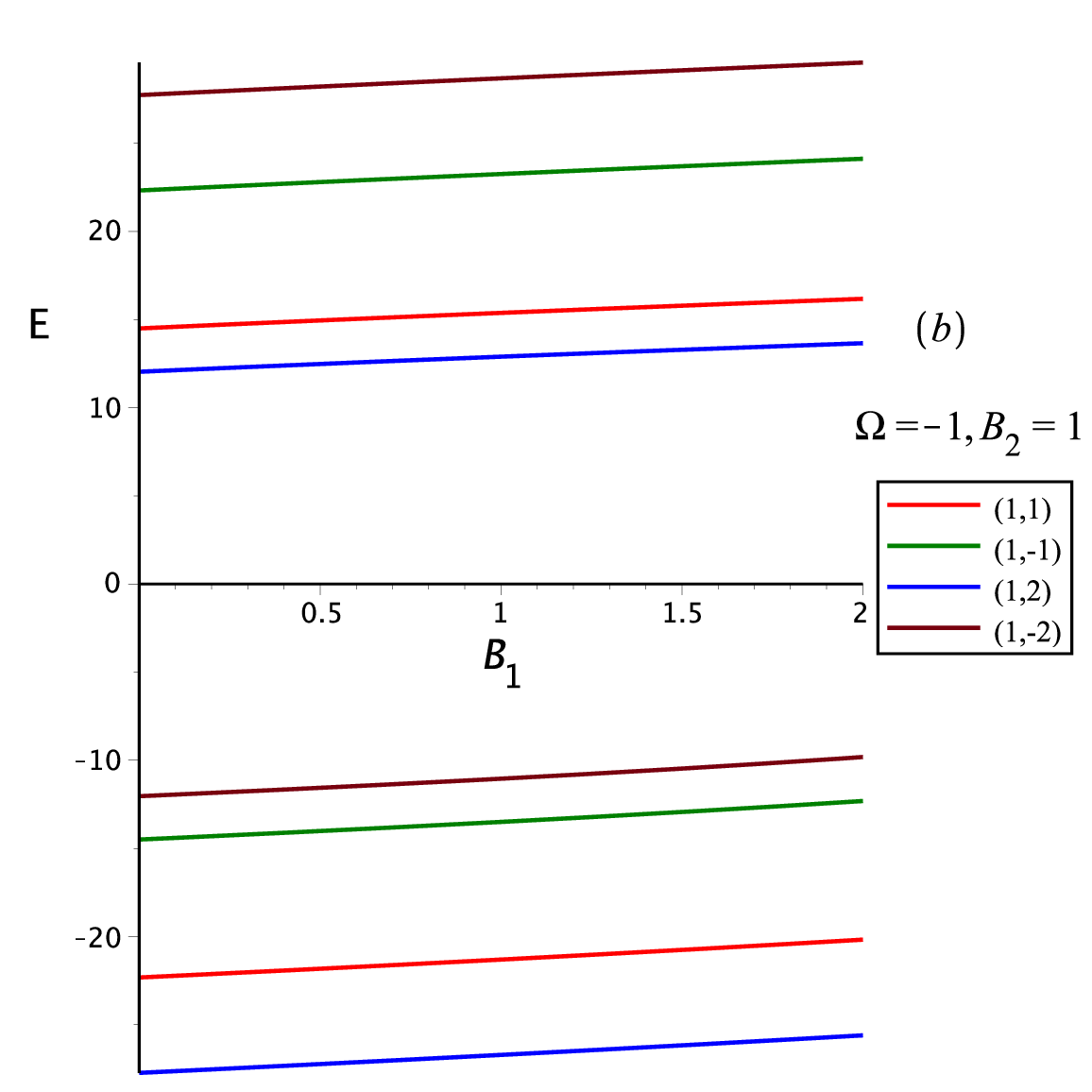}
\includegraphics[width=0.35\textwidth]{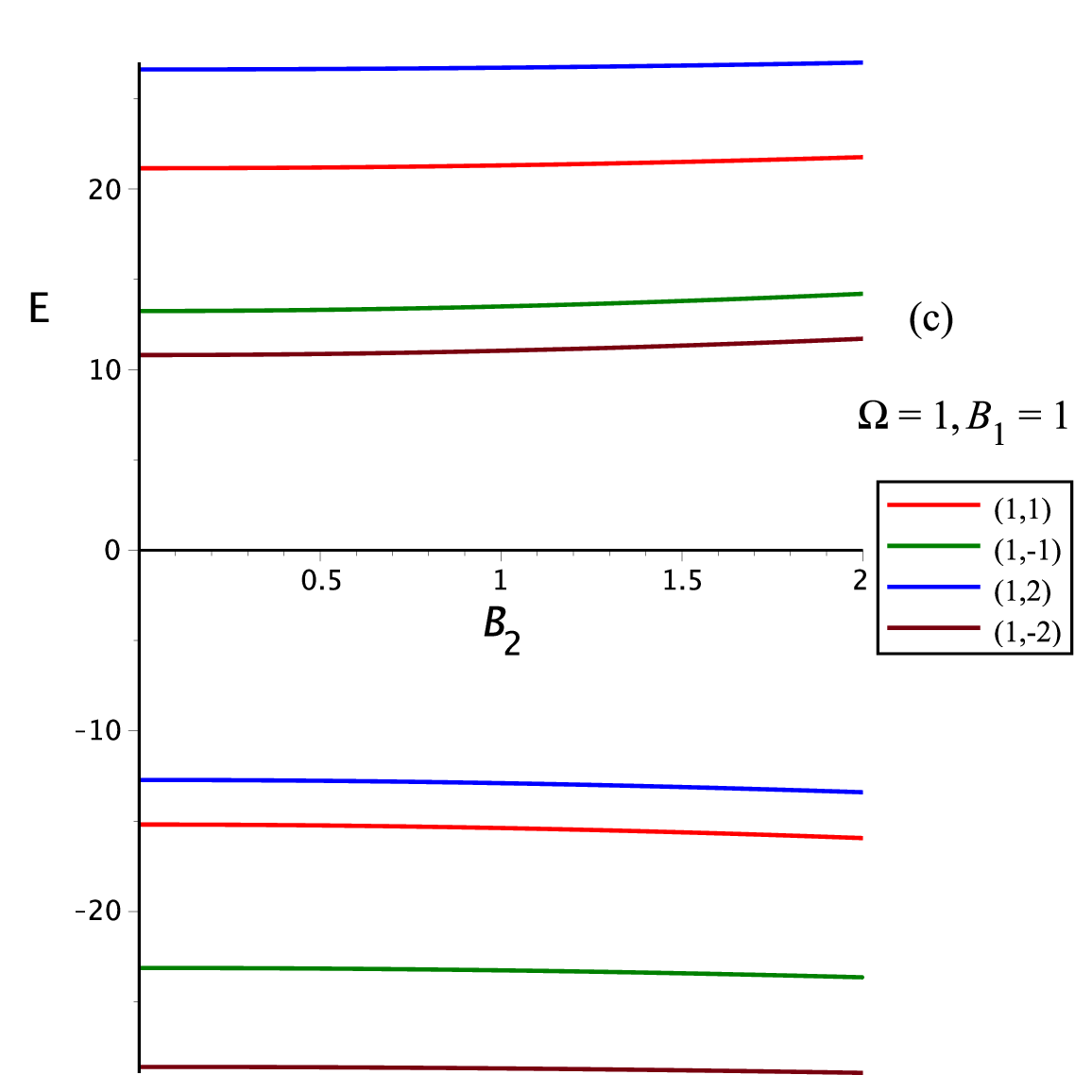}
\includegraphics[width=0.35\textwidth]{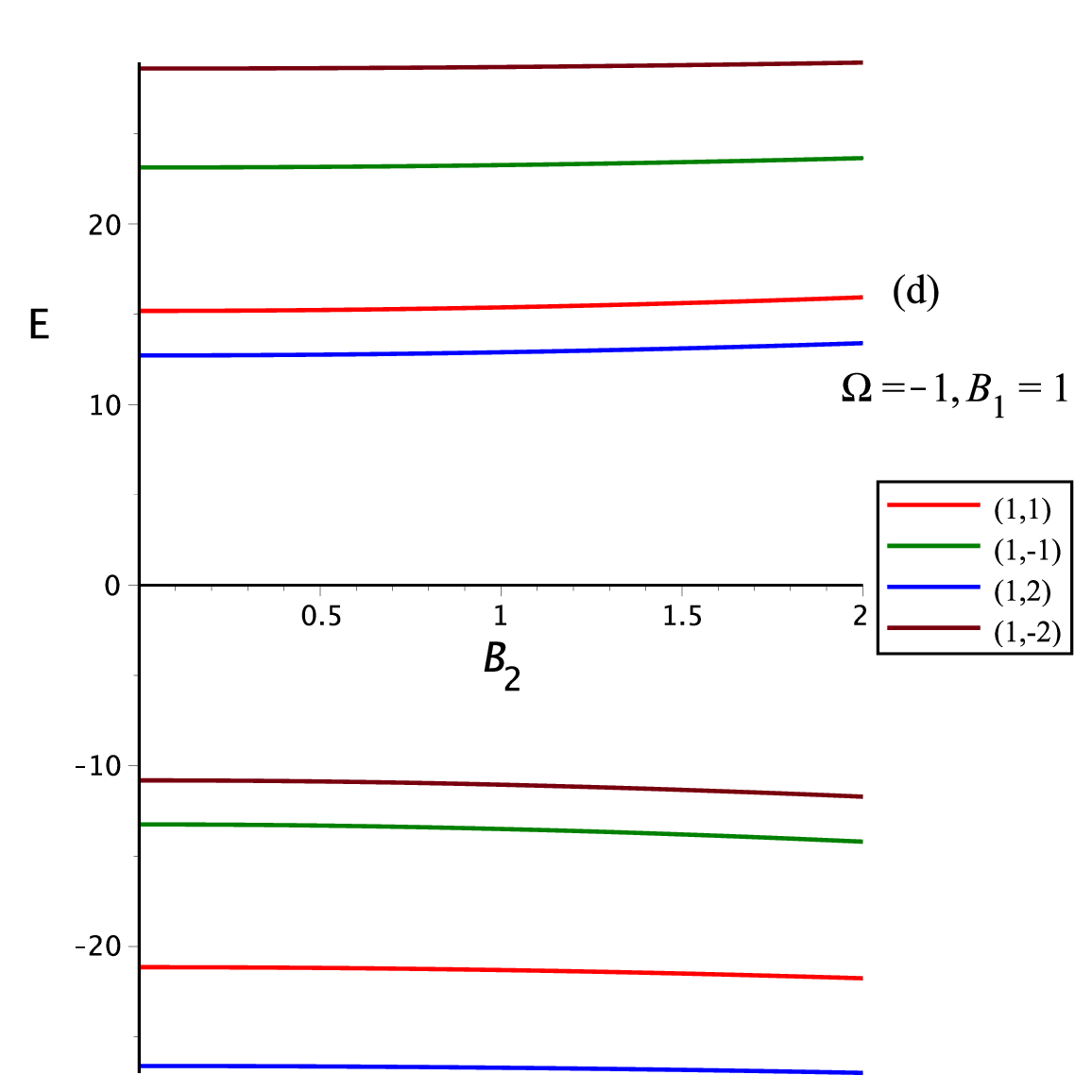}
\caption{\small 
{The energy levels against the magnetic fields, for $\left(
n_{r},m\right) $ states with $n_{r}=1$ and $m=\pm 1,\pm 2$ given by Eq. (\ref%
{2.97}), at $\alpha =0.5$, and $m_{\circ }=1=\eta =k$, so that 2(a) $E$
against $B_{1}$ for $\Omega =1$ and $B_{2}=1$, 2(b) $E$ against $B_{1}$ for $%
\Omega =-1$ and $B_{2}=1$, 2(c) $E$ against $B_{2}$ for $\Omega =1$ and $%
B_{1}=1$, and 2(d) $E$ against $B_{2}$ for $\Omega =-1$ and $B_{2}=1$.}}
\label{fig2}
\end{figure}%

\subsection{Switching off the uniform magnetic field, $B_{1}=0$}

When $B_{1}$ is switched off, the problem reduces into that for an effective
potential 
\begin{equation}
V_{eff}\left( r\right) =\tilde{\Omega}^{2}r^{2}-2\Omega E\tilde{B}_{2}^{2}r+%
\frac{2\tilde{m}\tilde{B}_{2}}{r}  \label{2.981}
\end{equation}%
one obtains%
\begin{equation}
E^{2}-2\tilde{m}\Omega E-\frac{\tilde{n}\left( 2\tilde{n}+1\right) }{%
\left\vert \tilde{m}\right\vert }\left\vert \Omega E\right\vert -\mathcal{G}%
_{n_{r},m}=0.  \label{2.99}
\end{equation}%
This result should be dealt with diligently since $\left\vert \Omega
E\right\vert =\Omega _{\pm }E_{\pm }$ (where $E=E_{\pm }=\pm |E|$ and $%
\Omega =\Omega _{\pm }=\pm |\Omega |$) or $\left\vert \Omega E\right\vert
=-\Omega _{\mp }E_{\pm }$. We have, therefore, two quadratic equations at
our disposal to handle. That is, for $\left\vert \Omega E\right\vert =\Omega
_{\pm }E_{\pm }$ we have 
\begin{equation}
E_{\pm }^{2}-\left[ 2\tilde{m}+\frac{\tilde{n}\left( 2\tilde{n}+1\right) }{%
\left\vert \tilde{m}\right\vert }\right] \Omega _{\pm }E_{\pm }-\mathcal{G}%
_{n_{r},m}=0,  \label{2.100}
\end{equation}%
and for $\left\vert \Omega E\right\vert =-\Omega _{\mp }E_{\pm }$ we have%
\begin{equation}
E_{\pm }^{2}-\left[ 2\tilde{m}-\frac{\tilde{n}\left( 2\tilde{n}+1\right) }{%
\left\vert \tilde{m}\right\vert }\right] \Omega _{\mp }E_{\pm }-\mathcal{G}%
_{n_{r},m}=0.  \label{2.101}
\end{equation}%
Under such settings, we obtain%
\begin{equation}
E_{\pm }=\pm \left[ \tilde{m}+\frac{\tilde{n}\left( \tilde{n}+1/2\right) }{%
\left\vert \tilde{m}\right\vert }\right] \,\left\vert \Omega \right\vert \pm 
\sqrt{\left[ \tilde{m}+\frac{\tilde{n}\left( \tilde{n}+1/2\right) }{%
\left\vert \tilde{m}\right\vert }\right] ^{2}\Omega ^{2}+\mathcal{G}%
_{n_{r},m}}  \label{2.102}
\end{equation}%
for $\left\vert \Omega E\right\vert =\Omega _{\pm }E_{\pm }$, and 
\begin{equation}
E_{\pm }=\mp \left[ \tilde{m}-\frac{\tilde{n}\left( \tilde{n}+1/2\right) }{%
\left\vert \tilde{m}\right\vert }\right] \,\left\vert \Omega \right\vert \pm 
\sqrt{\left[ \tilde{m}-\frac{\tilde{n}\left( \tilde{n}+1/2\right) }{%
\left\vert \tilde{m}\right\vert }\right] ^{2}\Omega ^{2}+\mathcal{G}%
_{n_{r},m}}  \label{2.103}
\end{equation}%
for $\left\vert \Omega E\right\vert =-\Omega _{\mp }E_{\pm }$. That is, $%
E_{\pm }$ in (\ref{2.102}) should exactly read%
\begin{equation}
E_{+}=+\left[ \tilde{m}+\frac{\tilde{n}\left( \tilde{n}+1/2\right) }{%
\left\vert \tilde{m}\right\vert }\right] \,\left\vert \Omega \right\vert +%
\sqrt{\left[ \tilde{m}+\frac{\tilde{n}\left( \tilde{n}+1/2\right) }{%
\left\vert \tilde{m}\right\vert }\right] ^{2}\Omega ^{2}+\mathcal{G}%
_{n_{r},m}},  \label{2.104}
\end{equation}%
and%
\begin{equation}
E_{-}=-\left[ \tilde{m}+\frac{\tilde{n}\left( \tilde{n}+1/2\right) }{%
\left\vert \tilde{m}\right\vert }\right] \,\left\vert \Omega \right\vert -%
\sqrt{\left[ \tilde{m}+\frac{\tilde{n}\left( \tilde{n}+1/2\right) }{%
\left\vert \tilde{m}\right\vert }\right] ^{2}\Omega ^{2}+\mathcal{G}%
_{n_{r},m}}.  \label{2.105}
\end{equation}%
Similarly, we obtain $E_{+}$ and $E_{-}$ for (\ref{2.103}) as%
\begin{equation}
E_{+}=-\left[ \tilde{m}-\frac{\tilde{n}\left( \tilde{n}+1/2\right) }{%
\left\vert \tilde{m}\right\vert }\right] \,\left\vert \Omega \right\vert +%
\sqrt{\left[ \tilde{m}-\frac{\tilde{n}\left( \tilde{n}+1/2\right) }{%
\left\vert \tilde{m}\right\vert }\right] ^{2}\Omega ^{2}+\mathcal{G}%
_{n_{r},m}},  \label{2.106}
\end{equation}%
and%
\begin{equation}
E_{-}=+\left[ \tilde{m}-\frac{\tilde{n}\left( \tilde{n}+1/2\right) }{%
\left\vert \tilde{m}\right\vert }\right] \,\left\vert \Omega \right\vert -%
\sqrt{\left[ \tilde{m}-\frac{\tilde{n}\left( \tilde{n}+1/2\right) }{%
\left\vert \tilde{m}\right\vert }\right] ^{2}\Omega ^{2}+\mathcal{G}%
_{n_{r},m}}.  \label{2.107}
\end{equation}%
One may clearly observe, without doubt, the symmetry of the energies about $%
E=0$ value for the set $\left\vert \Omega E\right\vert =\Omega _{\pm }E_{\pm
}$ in (\ref{2.104}) and (\ref{2.105}), and the set $\left\vert \Omega
E\right\vert =-\Omega _{\mp }E_{\pm }$ in  (\ref{2.106}) and (\ref{2.107}),
respectively.

\section{Concluding remarks}

In this work, we have studied the effects Som-Raychaudhuri rotating cosmic
string spacetime on the KG- oscillators in a mixed magnetic field. We have
observed that the corresponding KG-equation takes a Schr\"{o}dinger-like
form with an interaction potential in the form of 
\begin{equation}
V\left( r\right) =\tilde{\Omega}^{2}r^{2}+Ar-\frac{2\tilde{m}\tilde{B}_{2}}{r%
},  \label{3.1}
\end{equation}%
(i.e., it includes an oscillator, a linear, and a Coulomb-like
interactions). Such a Schr\"{o}dinger-like equation is shown to admit a
solution in the form of biconfluent Heun functions/series $H_{B}\left( 
\breve{\alpha},\beta ,\gamma ,\delta ,z\right) $, where the usual power
series expansion is truncated to a polynomial of \ order $n_{r}+1=n\geq 1$
through the usual condition $\gamma =2\left( n_{r}+1\right) +\breve{\alpha}$%
. We have also used the very recently developed parametric correlation $%
\delta =-\beta \left( 2n_{r}+\breve{\alpha}+3\right) $ used by Mustafa \cite%
{1.29} and Mustafa et al. \cite{1.46} to obtain a conditionally exact
solution to the problem at hand. Such a parametric correlation identifies an
alternative condition, than that suggested by Ronveaux \cite{1.47}. \
Consequently, we were able to discuss and report the effects the mixed
magnetic fields and the vorticity of the Som-Raychaudhuri rotating cosmic
string spacetime on the KG-oscillators spectroscopic structure (documented
in Fig.s 1 and 2).

Interestingly, we have observed that the competition between an effective
energy dependent vorticity (i.e., $\grave{\Omega}\left( E\right) =\Omega
E=\pm \left\vert \Omega \right\vert \left\vert E\right\vert $) and the
uniform magnetic field strength $\tilde{B}_{1}$ in (\ref{2.98.1}) plays a
crucial role in shaping the spectroscopic structure of the KG-oscillators
and breaks the symmetry of the corresponding energies about $E=0$ value.
However, when the uniform magnetic field is switched off (i.e., $B_{1}=0$)
such symmetry is retrieved (clearly documented in (\ref{2.103})) as such a
competition no longer exists. 

To the best of our knowledge, the current study has never been discussed
elsewhere in the literature. Yet, the current methodical proposal provide a
gateway to the study the thermodynamic properties of several quantum systems
described in non-trivial spacetime backgrounds, like internal energy,
entropy, specific heat, etc \cite{1.49,1.50,1.51,1.52,1.53,1.54,1.55}, to
mention a few.


\begin{thebibliography}{99}

\bibitem{1.1} T. W. B. Kibble, J. Phys. A \textbf{9} (1976) 1387.

\bibitem{1.2} T. W. B. Kibble, arXiv:astro-ph/0410073v2 "Cosmic Strings
Rebon?" (2004).

\bibitem{1.3} T. W. B. Kibble, Phys. Rep. \textbf{67} (1980) 183.

\bibitem{1.4} A. Vilenkin, Phys. Rep. \textbf{121} (1985) 263.

\bibitem{1.5} A. Vilenkin, Phys. Rev. D \textbf{23} (1981) 852.

\bibitem{1.6} E. R. Bezerra de Mello, V. B. Bezerra, Y. V. Grats, Class.
Quant. Grav. \textbf{15} (1998) 1915.

\bibitem{1.7} C. R. Muniz, V. B. Bezerra, Ann. Phys. \textbf{340} (2014) 87.

\bibitem{1.8} M. V. Sazhin et al., Mon. Not. R. Astron. Soc. \textbf{376 }%
(2007) 1731.

\bibitem{1.9} V. B. Bezerra, V. M. Mostepanenko, R. M. TeixeiraFilho, Int.
J. Mod. Phys. D \textbf{11} (2002) 437.

\bibitem{1.10} V. A. de Lorenci et al., Class. Quant. Grav. \textbf{16}
(1999) 3047.

\bibitem{1.11} J. Audretsch, A. Economou, Phys. Rev. D \textbf{44} (1991)
980.

\bibitem{1.12} B. Jensen, H. Soleng, Phys. Rev. D \textbf{45} (1992) 3528.

\bibitem{1.13} K. Jusufi, Eur. Phys. J. C \textbf{76} (2016) 332.

\bibitem{1.14} S. Deser, R. Jackiw, G 't Hooft, Ann. Phys. \textbf{152}
(1984) 220.

\bibitem{1.15} K. G\"{o}del, Rev. Mod. Phys. \textbf{21} (1949) 447.

\bibitem{1.16} M. J. Rebou\c{c}as, J. Tiomno, Phys. Rev. D \textbf{28}
(1983) 1251.

\bibitem{1.17} M. M. Som, A. K. Raychaudhuri, Prc. R. Soc. London \textbf{%
A304} (1968) 81.

\bibitem{1.18} O. Mustafa, Eur. Phys. J. Plus, \textbf{138} (2023) 21.

\bibitem{1.181} G. Grignani, C. Lee, Ann. Phys. \textbf{196} (1989) 386.

\bibitem{1.182} G. Cl\.{e}ment, Ann. Phys. \textbf{201} (1990) 241.

\bibitem{1.183} E. \v{S}im\'{a}nek, Phys. Rev. D \textbf{78} (2008) 045014.

\bibitem{1.184} K. Jusufi, Eur. Phys. J. C \textbf{76} (2016) 332.

\bibitem{1.185} O. Mustafa, Eur. Phys. J. C \textbf{84} (2024) 362.

\bibitem{1.186} N. Drukker, B. Fiol, J. Sim\'{o}n, JCAP \textbf{0410} (2004)
012.

\bibitem{1.187} J. Carvalho, A. M. de M. Carvalho, C. Furtado, Eur. Phys. J.
C \textbf{74} (2014) 2935.

\bibitem{1.188} B. D. B. Figueiredo, I. D. Soares, J. Tiomno, Class. Quant.
Grav. \textbf{9} (1992) 1593.

\bibitem{1.189} Z. Wang, Z. Long, C. Long, M. Wu, Eur. Phys. J. Plus \textbf{%
130} (2015) 36.

\bibitem{1.1810} S. Das, G. Gegenberg, Gen. Rel. Grav. \textbf{40} (2008)
2115.

\bibitem{1.1811} J. Carvalho, A. M. de M. Carvalho, E. Cavalcante,C.
Furtado, Eur. Phys. J. C \textbf{76} (2016) 365.

\bibitem{1.1812} G. Q. Garcia, J. R. de S. Oliveira, K. Bakke, C. Furtado,
Eur. Phys. J. Plus, \textbf{132} (2017) 123.

\bibitem{1.1813} F. Ahmed, Eur. Phys. J. C \textbf{78} (2018) 598.

\bibitem{1.1814} B. C. L\"{u}tf\"{u}o\^{g}lu, J. K\v{r}\'{\i}\v{z}, P.
Sedaghatnia, H. Hassanabadi, Eur. Phys. J. Plus, \textbf{135} (2020) 691.

\bibitem{1.19} M. S. Conha, C. R. Muniz, H. R. Christiansen, V. B. Bezerra,
Eur. Phys. J. C \textbf{76} (2016) 512.

\bibitem{1.20} C. R. Muniz, V. B. Bezerra, M. S. Conha, Ann. Phys. \textbf{%
350} (2014) 105.

\bibitem{1.21} G. de A. Marques, C. Furtado, V. B. Bezerra, F. Moraes, J.
Phys. A: Math. Theor. \textbf{34} (2001) 5945.

\bibitem{1.22} K. Bakke. L. R. Ribeiro, C. Furtado, J. R. Nascimento, Phys
Rev. D \textbf{79} (2009) 024008.

\bibitem{1.23} K. Bakke, Braz. J. Phys. \textbf{42} (2012) 437.

\bibitem{1.24} E. R. F. Medeiros, E. R. B. de Mello, Eur. Phys. J. C \textbf{%
72} (2012) 2051.

\bibitem{1.25} O. Mustafa, Nucl. Phys. B \textbf{995} (2023) 116334.

\bibitem{1.26} O. Mustafa, Ann. Phys. \textbf{440} (2022) 168857.

\bibitem{1.27} O. Mustafa, Phys. Lett. B \textbf{839} (2023) 137793.

\bibitem{1.28} O. Mustafa, Phys. Lett. B \textbf{850} (2024) 138482.

\bibitem{1.29} O. Mustafa, Eur. Phys. J. C \textbf{84} (2024) 362.

\bibitem{1.30} M. J. Bueno, C. Furtado, A. M. de M. Caravalho, Eur. Phys. J.
B \textbf{85} (2012) 53.

\bibitem{1.31} M. Moshinsky, A Szczepaniak, J. Phys. A: math. Gen. \textbf{22%
} (1989) L817.

\bibitem{1.32} B. Mirza, M. Mohadesi, Commun. Theor. Phys. \textbf{42}
(2004) 664.

\bibitem{1.33} K. Bakke, H. F. Mota, Eur. Phys. J. Plus \textbf{133} (2018)
409.

\bibitem{1.34} J. Cravalho, C. Furtado, F Moraes, Phys. Rev. A \textbf{84}
(2011) 032109.

\bibitem{1.35} O. Mustafa, Eur. Phys. J. C \textbf{82} (2022) 82.

\bibitem{1.36} O. Mustafa, Int. J. Geom. Meth. Mod. Phys. \textbf{20} (2023)
2350221.

\bibitem{1.37} K. Bakke, C. Furtado, Ann. Phys. \textbf{355} (2015) 48.

\bibitem{1.38} R. L. L. Vit\'{o}ria, H. Belich, Eur. Phys. J. C \textbf{78}
(2018) 999.

\bibitem{1.39} R. L. L. Vit\'{o}ria, K. Bakke, Eur. Phys. J. Plus \textbf{133%
} (2018) 490.

\bibitem{1.40} R. L. L. Vit\'{o}ria, H. Belich, K. Bakke, Eur. Phys. J. Plus 
\textbf{132} (2017) 25.

\bibitem{1.41} J. Cravalho, A. M. Cravalho, E. Cavalcante, C. Furtado, Eur.
Phys. J. C \textbf{76} (2016) 365.

\bibitem{1.42} H. Hassanabadi, M. Hosseinpour, Eur. Phys. J. C \textbf{76}
(2016) 553.

\bibitem{1.43} C. F. S. Rereira, A. R. Soares, R. L. L. Vit\'{o}ria, H.
Belich, Eur. Phys. J. C \textbf{83} (2023) 270.

\bibitem{1.44} A. R. Soares, R. L. L. Vit\'{o}ria, C. F. S. Rereira, Eur.
Phys. J. C \textbf{83} (2023) 903.

\bibitem{1.45} C. F. S. Rereira, R. L. L. Vit\'{o}ria, A. R. Soares, H.
Belich, Int. J. Theor. Phys. \textbf{62} (2023) 225.

\bibitem{1.46} O. Mustafa, A. R. Soares, C. F. S. Pereira, R. L. L. Vit\'{o}%
ria, Eur. Phys. J. C \textbf{84} (2024) 405.

\bibitem{1.47} A. Ronveaux, \textit{Heun's Differential Equations} (Oxford University Press, New York, 1995).

\bibitem{1.48} T A Ishkhanyan, V P Krainov, A M Ishkhanyan, J. Phys.: Conf.
Series \textbf{1416} (2019) 012014.

\bibitem{1.49} H. Hassanabadi, M. Hosseinpour, Eur. Phys. J. C \textbf{76}
(2016) 553.

\bibitem{1.50} A. N. Ikot, B. C. Lutfuoglu, M. I. Ngwueke, M. E. Udoh, S.
Zare, H. Hassanabadi, Eur. Phys. J. Plus \textbf{131} (2016) 419.

\bibitem{1.51} M. Eshghi, H. Mehraban, Eur. Phys. J. Plus \textbf{132}
(2017) 121.

\bibitem{1.52} B. Khosropour, Indian. J. Phys. \textbf{92} (2018) 43.

\bibitem{1.53} A. N. Ikot, U. S. Okorie, R. Sever, and G. J. Rampho, Eur.
Phys. J. Plus \textbf{134} (2019) 386.

\bibitem{1.54} R. L. L. Vit\'oria, H. Belich, Adv.
High Energy Phys. \textbf{2020} (2020) 4208161.

\bibitem{1.55} R. L. L. Vit\'oria, Tiago Moy, H. Belich, Few-Body Syst. \textbf{63} (2022) 51.
\end{thebibliography}
\end{document}